\documentclass[referee]{aa} 
\usepackage{graphicx}
\begin{document}

\title{Radial velocity measurements of B stars in the Scorpius-Centaurus association}
\author{E.\,Jilinski\inst{1,2}, S. Daflon\inst{1}, K.\,Cunha \inst{1} and
R.\,de la Reza \inst{1} }

\offprints{E.\,Jilinski, Observat\'orio Nacional/MCT, Rua Gal. Jose
   Cristino 77, S\~ao Cristov\~ao, Rio de Janeiro, Brazil.
   e-mail: jilinski@on.br}

\institute{Observat\'orio Nacional/MCT, Rio de Janeiro, Brazil
\and
Main Astronomical Observatory, Pulkovo, St. Petersburg, Russia}

\date{}

\authorrunning{E.\,Jilinski et al.}

\titlerunning{Radial velocities in the Sco-Cen association}

\abstract{We derive single-epoch radial velocities for a sample of
56 B-type stars members of the subgroups Upper Scorpius, Upper
Centaurus Lupus and Lower Centaurus Crux of the nearby Sco-Cen OB
association. The radial velocity measurements were obtained by
means of high-resolution echelle spectra via analysis of
individual lines. The internal accuracy obtained in the
measurements is estimated to be typically 2-3 km\,s$^{-1}$, but
depends on the projected rotational velocity of the target. Radial
velocity measurements taken for 2-3 epochs for the targets
HD120307, HD142990 and HD139365 are variable and confirm that they
are spectroscopic binaries, as previously identified in the
literature. Spectral lines from two stellar components are
resolved in the observed spectra of target stars HD133242,
HD133955 and HD143018, identifying them as spectroscopic binaries.
\keywords{stars: early-type - stars: binaries: spectroscopic -
stars:
          kinematics - stars: radial velocities in open clusters and associations:
          individual:  Scorpius-Centaurus association. }
 }

\maketitle

\section{Introduction}

The Scorpius-Centaurus association is the nearest association of
young OB stars to the Sun. Blaauw (1960, 1964) divided this
association into three stellar subgroups: Upper Scorpius (US),
Upper Centaurus Lupus (UCL) and Lower Centaurus Crux (LCC). LCC
and UCL have roughly similar ages of about $16-20$ Myr, while US
is younger with an estimated age of $\sim$ 5 Myr (Mamajek et
al.2002; Sartori et al. 2003). This complex OB association of
unbound stars is of great interest because, as recently shown, it
is related to the origins of nearby  moving groups of low mass
post-T Tauri stars with ages around 10 Myr: the $\beta$ Pictoris
Moving Group, the TW Hydra association, and the $\eta$ and
$\epsilon$ Chamaleonis groups (Mamajek et al. 2000; Ortega et al.
2000, 2004; Jilinski et al. 2005). In addition, the
Scorpius-Centaurus association also appears to be the source of a
large bubble of hot gas in which the Sun is plunged. All these
structures are believed to have been possibly triggered by
supernova explosions taking place in UCL and LCC during the last
~13 Myr (Ma\'iz-Apell\'aniz 2001).

The technique adopted for investigating the origins of the $\beta$
Pictoris Moving Group, for example, consists of tracing back the
3-D stellar orbits of the members of these moving groups until
their main first orbits confinement was found, as well as the past
mean positions of LCC and UCL. This enabled, investigator not only
to determine the dynamical age of this moving group, but also to
investigate properties of their birth clouds (Ortega et al. 2002,
2004). It is also possible to find the past positions of the
possible supernovae that triggered the formation of these groups
by tracing back the orbit of a runaway OB star, which could have
been the result of a supernova explosion in LCC or UCL (see, for
example, Hoogerwerf et al. 2001 and Vlemmings et al. 2004).

While the past evolution of these moving groups of low mass stars
appears to be a relatively simple problem  (as the dynamical ages
are not so old), the dynamical evolution of the older and more
numerous subgroups LCC and UCL appears to be more difficult. There
is the possibility of the presence of several generations of hot
stars during the mainstream of the OB association evolution
(Garmany 1994).
   Substructure in LCC and UCL was found by de Bruijne (1999), based
on Hipparcos data. The formation of the younger US subgroup could
have been triggered by UCL some 6-8 Myr ago (Preibisch et al.
2001). All these studies require reliable radial velocities in
order to calculate space velocities. In this paper we present
single-epoch radial velocity (RV) measurements for 56 B-type stars
members of LCC, UCL and US subgroups, to contribute to studies of
their dynamics so as to unravel their origins.

\section{Observations and reduction}

A sample of 56 B-type stars from the Scorpius-Centaurus
association was observed during observing runs in May 16-20 and
July 7, 2002, with the 1.52m telescope equipped with the FEROS
echelle spectrograph (Kaufer et al.  2000; resolving power
R=48,000, wavelength coverage between 3900 and 9200\AA.) with a
CCD detector at the European Southern Observatory (ESO) \footnote
{Observations obtained under the ON/ESO agreement}. The target
stars were selected from the list in Humphreys \& McElroy (1984)
and from the comprehensive study of OB associations based on
Hipparcos observations by de Zeeuw et al. (1999). The observed
targets are listed in Table~1. From this sample, according to de
Zeeuw et al. (1999), 15 targets are confirmed members of the LCC,
while 15  stars are members of the UCL and 11 stars are from the
US subgroup. For the remaining 15 stars in our sample, membership
to any of these subgroups was not certain.

The spectra were reduced with the MIDAS reduction package and
consisted of the following standard steps: CCD bias correction,
flat-fielding, extraction, wavelength calibration, correction of
barycentric velocity, as well as spectrum rectification and
normalization.  The one-dimensional spectra were then treated by
tasks in the NOAO/IRAF data package. The signal-to noise ratio
obtained in the observed spectra was typically larger than 100 and
typical exposure times varied between 300 seconds for the
brightest stars (V $\sim$ 3) and 1200 seconds for stars with V
$\sim$ 5. Typical spectra are shown in Figure~1 . The top panel
corresponds to the target star HD~122980 and the bottom panel to
HD~112092. Both stars have sharp lines with projected rotational
velocities ($v \sin i$) less than ~40 km\,s$^{-1}$. The spectral
region displayed shows identifications of several lines that were
used in the RV determinations. The FEROS bench spectrograph and
set up have proven to have high spectral stability for RV
measurements as concluded from a study of radial-velocity standard
stars: a r.m.s. of 21 m\,s$^{-1}$ has been obtained for a data set
of 130 individual measurements (Kaufer et al. 2000).

\section{Analysis and discussion}

The cross-correlation technique, which is used for precise RV
determinations in later type stars, when applied to the hotter OB
stars can be problematic as early type star spectra show few
absorption lines. These lines are in many cases, intrinsically
broad (up to a few hundreds km\,s$^{-1}$) due to stellar rotation.
In addition, there is also the possibility of line variability
affecting their line profiles (Steenbrugge et al. 2003).
Therefore, the cross-correlation peak that defines the value of
radial velocity can be very broad and contain important
sub-structures caused by blending of spectral lines that appear to
have different widths. In addition to having high $v \sin i$
values many OB stars are binary and it is not straightforward to
apply the cross-correlation method and to identify them as
double-lined binaries; in order to obtain the orbital solution a
long set of observations is needed. Detailed cross-correlation
technique analyses applied to determinations of radial velocities
of early-type stars has been presented in a number of recent
publications (see, for example, Verschueren et al. 1997;
Verschueren et al. 1999a; Griffin et al. 2000). Griffin et al.
(2000), in particular, discuss in detail the difficulties in
obtaining accurate RV measurements from cross-correlation in
early-type stars spectra.

In this study, having high-resolution observations covering a
large spectral range, radial velocity values for the target stars
were obtained from measurements of the positions of individual
spectral lines of He I, C II, N II, O II, Mg II, Si II and Si III,
relative to their rest wavelengths. Radial-velocity standard stars
were not observed. (The adopted linelists can be found in Daflon
et al. 2001, 2003.) We inspected and identified all unblended
lines visible in the spectral range between 3798 \AA\ (H\,{\sc i})
and 7065 \AA\ (He\,{\sc i}) in each target star: the number of
measurable lines varied between 10 and 74, depending on the star
spectral type, rotation velocity, possible multiplicity, but also
on the signal-to-noise of the obtained spectra. (We note, however,
that for the double lined binary HD 133242, it was possible to
measure positions only for 4 lines in component A and 6 lines in
component B.) Mean radial velocities using all measurable lines
(${\rm RV}$) and respective dispersions were calculated for the
individual target stars.

In Table~1 we assemble our RV results as well as results from the
literature. In the two first columns of this table we list the HD
numbers of the observed stars with the respective spectral types;
in the columns 3 and 4 we list the heliocentric Julian Date (HJD)
and the measured radial velocities, plus the number of measured
lines in brackets. In the other columns we list results from the
literature: columns 5 and 6 list the RV$_{\rm GCRV}$ and
associated error or quality, from the General Catalogue of Radial
Velocities (GCRV; quality flags A to E, or I for insufficient
data); column 7 presents the projected rotational velocity from
Brown \& Verschueren (1997) and, when not available in this
source, the $v\,sin i$ was taken from the compilation of Glebocki
\& Stawikowski (2000); in column 9 we list, when available,
literature references where information about duplicity can be
found for the stars. The stars are separated according to the
different subgroups in the Scorpius-Centaurus association,
following the membership probabilities P(m) listed by de Zeeuw et
al. (1999).

The internal precision of our RV determinations can be represented
by the scatter obtained from the RV measurements line-by-line,
which is listed in column 4 of Table 1. These are typically
smaller than $\sim$2.0 ${\rm km\,s}^{-1}$ for stars with estimated
$v\,sin i$ smaller than 100 ${\rm km\,s}^{-1}$. We note, however,
that when the target $v\,sin i$ are large, the uncertainties in
the derived RVs can be significantly larger due to uncertainties
in defining the line center. This can be seen in Figure~2, where
we show the obtained line-to-line scatter versus the target
projected rotational velocity (as taken from Brown \& Verschueren
1997 and Glebocki \& Stawikowski 2000).

In order to evaluate possible systematic effects that line
selection could have on the RV results, we selected a  homogeneous
set of 28 spectral lines of  H\,{\sc i}, He\,{\sc i}, Si\,{\sc
iii} and Mg\,{\sc ii}, that could be measured in most of the
studied spectra, and recalculated the mean radial velocity values
for all possible stars. A comparison of the mean radial velocities
${\rm RV}$ with ${\rm RV}_{\rm 28lines}$ (obtained using only the
selected 28 lines) indicates that there are non-significant
systematic differences between the two determinations ${\rm RV} -
\, {\rm RV}_{\rm 28lines}= -1.0 {\rm km\,s}^{-1}$, with $\sigma  =
3.1 {\rm km\,s}^{-1}$.

\subsection{Membership}

Table~1 lists the target stars according to their membership the 3
subgroups as assigned by de Zeeuw et al. (1999). Most of the stars
in the Lower Centaurus Crux subgroup are flagged as binaries in
the literature, except for HD103079, HD106490 and HD108483. For
these 3 stars we measured radial velocities of $19.3 {\rm
km\,s}^{-1}$, $15.3 {\rm km\,s}^{-1}$ and $12.8 {\rm km\,s}^{-1}$,
respectively, with RV$_{mean}$=15.8 $\pm$3.2 ${\rm km\,s}^{-1}$.
This mean value is in general agreement with the mean radial
velocity calculated by de Zeeuw et al. (1999) for LCC, which is of
$12 {\rm km\,s}^{-1}$. For the subgroup UCL, our sample has 2
non-binary stars (HD121790 and HD128345) and  RV$_{mean}$=$9.6
{\rm km\,s}^{-1}$ which is $\sim 5 {\rm km\,s}^{-1}$ higher than
the de Zeeuw et al. (1999) mean value of $4.9 {\rm km\,s}^{-1}$.
For the Upper Scorpius subgroup all the sample stars have been
flagged as binaries in the literature.

For the five target stars that had not been identified as members
of any of the three subgroups in the Sco-Cen association (listed
as "others" in Table~1) and for which we have no information on
duplicity, we can attempt to discuss their membership status based
on the comparison of the radial velocities measured here and in
the literature. We find that the measured radial velocities for
HD109026 (RV=4.0 ${\rm km\,s}^{-1}$) and HD110335 (RV= 4.8 ${\rm
km\,s}^{-1}$) are consistent with the mean radial velocity for UCL
of 4.9 ${\rm km\,s}^{-1}$s. For the target star HD109026 we have
RV$_{GCRV}$= 2.5 ${\rm km\,s}^{-1}$, therefore it could be
considered initially as having constant RV (within the
uncertainties) and possibly a member of the Upper Centaurus Lupus
subgroup. For HD110335, we find a larger discrepancy between our
measurement (RV=4.8 ${\rm km\,s}^{-1}$) and the RV value in the
GCRV (RV$_{GCRV}$=12.5 ${\rm km\,s}^{-1}$). The RVs,  however, are
marginally consistent given the expected uncertainty brackets that
affect the 2 determinations. If this is really the case, HD110335
could be also considered as a possible member of the UCL subgroup.
In addition, the target star HD120640 (RV$_{mean}$=-1.8 ${\rm
km\,s}^{-1}$ from this study and RV$_{GCRV}$=-4.7 ${\rm
km\,s}^{-1}$) can be assumed here to have constant radial
velocity. These measurements are consistent with the mean RV value
of -4.6 ${\rm km\,s}^{-1}$ listed by de Zeeuw et al. (1999) for
this subgroup.

The two other stars in our sample of 'others' (HD115846 and
HD105937) for which we derived RV= -21.8 ${\rm km\,s}^{-1}$ and
RV=22.7 ${\rm km\,s}^{-1}$, respectively, have values in the GCRV
of RV$_{GCRV}$=3.0 ${\rm km\,s}^{-1}$ and RV$_{GCRV}$=15.0 ${\rm
km\,s}^{-1}$. We found no information in the literature about
these stars  being confirmed binary stars, but the variation in RV
for HD115846 exceeds the expected uncertainties: this target
probably has a non-constant RV, which prevents further
considerations about it belonging to any of the Sco-Cen subgroups.
HD105937 has an RV only marginally constant within the
uncertainties, but its mean RV is not compatible with any of the
subgroups.

\subsection{Duplicity}

Results from a search for duplicity information for the targets
stars (column 9; Table~1) indicate that a large number of stars in
our sample are flagged as binaries in the literature. For most of
these targets we have only one single-epoch RV measurement and our
results alone cannot be used to infer duplicity. However, the RVs
derived in this study can be added to RV databases and contribute
to long term studies of their orbits. Only a small number of stars
had not been previously flagged as binaries in the different
studies in the literature. For this subsample of 10 stars,
considered a priori as RV constants, it is possible to compare our
RV values with the averaged radial velocities assembled in the
GCRV. This comparison is shown in Figure~3. Our RV determinations
compare favorably with the RVs from the GCRV with a scatter of the
order of the estimated uncertainties. ${\rm RV}_{\rm GCRV} - {\rm
RV} = 0.7$ and $\sigma = 4.9 {\rm km\,s}^{-1}$. (This was
calculated excluding one discrepant star, HD115846, which could be
a binary system.) Taking into account the mean precision of RV
determinations from the GCRV as $\pm  5 {\rm km\,s}^{-1}$, the
external precision of our RV determinations may be evaluated as
approximately $\pm  5 {\rm km\,s}^{-1}$.

For those targets with more than one epoch RV measurement in our
study, a subsample showed radial velocity variations larger than
the expected uncertainties: HD120307, HD142990 and HD139365. Since
these had been previously identified as SBs in the literature
(Levato et al. 1987 and Batten et al. 1989), our results confirm
their duplicity. Four other stars with multiple epoch observations
in this study showed a constant RV within the uncertainties:
HD116087, HD130807, HD132200 and HD120640.

For 3 targets in our sample (HD133242, HD133955 and HD143018) we
were able to separate and identify lines of two stellar
components, classifying them as double lined spectroscopic
binaries. Their combined spectra showing spectral lines from two
stars are shown in Fig~4. Two of these stars (HD143018 and
HD133955) were previously identified in Batten et al. (1999) as a
spectroscopic binaries.

\begin{acknowledgements}
We thank the anonymous referee for suggestions that significantly
improved the paper. E.G.J. thanks FAPERJ and MCT Brazil for
financial support.
\end{acknowledgements}

{}


\newpage

\begin{table}
\begin{center}
\begin{tabular}{llcrcccc}
\multicolumn{8}{c}{Table 1. Radial velocities of observed Sco-Cen OB stars}\\
\hline
\hline
 \,\,\, HD &\,\,\,\,Sp& HJD  &RV\,[N]\,\,\,\,\,\,\,\,& RV$_{\rm GCRV}$\,\,\, & eRV* & V$\sin$ i & Duplicity \\
   & & & (km\,s$^{-1}$)\,\,\,\,\,\, & (km\,s$^{-1}$) & (km\,s$^{-1}$) & (km\,s$^{-1}$) &  \\
\hline
\multicolumn{8}{c}{Lower Centaurus Crux} \\
\,\,\,98718&B5Vn  & 52411.0285 &   26.8$\pm$2.7 [19] &    9.4 &     C &  340$^a$ & ST \\
 103079  & B4V    & 52411.0425 &   19.3$\pm$1.0 [58] &   20.6 &     B &   47$^b$  &      \\
 105382  & B6IIIe & 52413.0429 &   15.8$\pm$1.1 [50] &   16.4 &     B &   75$^b$  &  ST \\
 106490  & B2IV   & 52413.0617 &   15.3$\pm$1.5 [35] &   22.0 &     B &  135$^b$  &      \\
 106983  & B2.5V  & 52411.0787 &   12.1$\pm$0.8 [50] &   15.8 &     A &   65$^b$  &  VDH \\
 108257  & B3Vn   & 52411.0875 &   19.9$\pm$2.7 [19] &    5.0 &     C &  298$^b$  &  VDH \\
 108483  & B2V    & 52411.1008 &   12.8$\pm$1.4 [29] &    8.0 &     C &  169$^b$  &      \\
 109668  & B2IV-V & 52411.1127 & $-$0.1$\pm$1.4 [38] &   13.0 &     C &  114$^b$  &  ST \\
 110879  & B2.5V  & 52411.1177 &   56.9$\pm$2.5 [33] &   42.0 &     D &  139$^b$  &  VDH \\
 110956  & B3V    & 52411.1229 &   15.6$\pm$0.7 [69] &   16.4 &     B &   26$^b$  &  VDH \\
 112091  & B5Vne  & 52411.1449 &   15.9$\pm$1.9 [19] &   13.0 &     C &  242$^b$  &VDH\\
 112092  & B2IV-V & 52411.1389 &   14.4$\pm$0.6 [72] &   13.9 &     A &   34$^b$  &  VDH \\
 113703  & B5V    & 52415.1623 &    1.2$\pm$1.4 [29] &    6.0 &     C &  140$^b$  &  VDH,ST \\
 113791  & B1.5V  & 52411.1615 &   58.5$\pm$0.8 [72] &   14.3 &     C &   15$^b$  &  SB8 \\
 116087  & B3V    & 52411.2216 &   12.3$\pm$1.6 [21] &    6.0 &     C &  233$^b$  &  VDH,ST \\
         &        & 52414.1034 &    9.4$\pm$3.9 [10] &        &       &       &      \\
\multicolumn{8}{c}{Upper Centaurus Lupus} \\
 120307  & B2IV   & 52413.1160 &   25.4$\pm$0.9 [54] &    9.1 &       &   65$^b$  & SB8,L87\\
         &        & 52413.1562 &    7.4$\pm$0.8 [39] &        &       &       &      \\
         &        & 52414.1143 &   11.0$\pm$0.7 [31] &        &       &       &      \\
 121743  & B2IV   & 52414.1453 &    9.6$\pm$0.8 [57] &    5.3 &   1.4 &   79$^b$  & L87\\
 121790  & B2IV-V & 52414.1523 &    9.2$\pm$1.4 [35] &    4.8 &     B &  124$^b$  &      \\
 122980  & B2V    & 52414.1594 &   10.5$\pm$0.6 [74] &    9.6 &   2.8 &   15$^b$  & L87  \\
 128345  & B5V    & 52414.1828 &    9.9$\pm$1.9 [25] &    8.0 &     D &  186$^b$  &      \\
 129056  & B1.5III& 52414.1909 &   18.3$\pm$0.8 [72] &    5.4 &   0.6 &   16$^b$  & VDH\\
 130807  & B5IV   & 52414.1957 &    7.1$\pm$0.6 [67] &    7.3 &     A &   27$^b$  & VDH,ST\\
         &        & 52414.2150 &    7.2$\pm$0.5 [56] &        &       &       &      \\
 132200  & B2IV   & 52414.2097 &    4.6$\pm$0.6 [70] &    8.0 &   0.9 &   32$^b$  &VDH,L87,ST\\
         &        & 52414.2266 &    4.9$\pm$0.6 [55] &        &       &       &      \\
133242A\,& B5IV   & 52414.2599 &$-$52.5$\pm$0.7 [4]  &    4.5 &    C  &  140$^a$ &  VDH\\
133242B\,&        &            &   81.5$\pm$2.1 [6]  &        &       &       &      \\
133955A\,& B3V    & 52414.2696 &   61.2$\pm$1.4 [10] &    9.8 &    B  &  135$^b$  & SB8\\
133955B\,&        &            &$-$31.6$\pm$1.0 [10] &        &       &       &      \\
 134687  & B3IV   & 52414.2777 &   23.7$\pm$0.8 [74] &   13.5 &    D  &   13$^b$  &  SB8 \\
 136504  & B2IV-V & 52414.2904 & $-$5.7$\pm$1.1 [60] &    7.9 &    C  &   41$^b$  &  SB8,ST \\
 137432  & B4Vp   & 52414.2976 &    6.3$\pm$1.0 [41] & $-$0.8 &    E  &   77$^b$  &  VDH,SB8 \\
 139365  & B2.5V  & 52414.1535 &   33.3$\pm$2.3 [23] &$-$14.0 &    E  &  134$^b$  &  SB8 \\
         &        & 52414.3140 & $-$5.8$\pm$1.1 [25] &        &       &       &      \\
 140008  & B5V    & 52414.3210 &    2.6$\pm$0.8 [56] &    3.9 &    B  &   11$^b$  &  SB8,ST \\
\hline
\end{tabular}
\end{center}
\end{table}

\begin{table}
\begin{center}
\begin{tabular}{llcrcccc}
\multicolumn{8}{c}{Table 1. Continued} \\
\hline
\hline
 \,\,\, HD &\,\,\,\,Sp& HJD  &RV\,[N]\,\,\,\,\,\,\,\,& RV$_{\rm GCRV}$\,\,\, & eRV* & V $\sin$ i & Duplicity \\
   & & & (km\,s$^{-1}$)\,\,\,\,\,\, & (km\,s$^{-1}$) & (km\,s$^{-1}$)  & (km\,s$^{-1}$) &  \\
\hline
\multicolumn{8}{c}{Upper Ssorpius} \\
 142669  & B2IV-V & 52415.2944 &    2.5$\pm$1.0 [41] &    3.3 &    E  &   98$^b$  &  SB8 \\
 142883  & B3V    & 52415.3002 &$-$54.3$\pm$0.5 [70] &$-$27.5 &    E  &   14$^b$  &  SB8 \\
 142990  & B5V    & 52481.0065 & $-$5.6$\pm$3.5 [33] &$-$12.1 &   3.4 &  178$^b$  & L87  \\
         &        & 52481.1270 &$-$10.9$\pm$2.9 [34] &        &       &       &      \\
 143018A\,& B1V+  & 52414.3559 &  113.8$\pm$2.8 [25] &$-$11.7 &    D  &  100$^b$  &  SB8 \\
 143018B\,&       &            &$-$173.6$\pm$3.2 [21]&        &       &       &      \\
 144217  & B0.5V  & 52414.3646 &    9.1$\pm$1.3 [43] & $-$1.0 &     & 91$^b$ &SB8,VDH,L87,ST\\
 144470  & B1V    & 52415.2873 & $-$0.6$\pm$1.1 [39] & $-$4.4 &   3.0 &  100$^b$  &  L87 \\
 147165  & B1III  & 52414.3498 &$-$25.6$\pm$1.2 [50] &    2.5 &       &   56$^b$  &SB8,L87,ST\\
 147888  & B3/B4V & 52481.1481 & $-$3.8$\pm$2.3 [27] & $-$6.8 &   2.9 &  175$^a$ &  L87 \\
 147932  & B5V    & 52481.2006 & $-$2.8$\pm$2.7 [23] &$-$11.0 &   2.4 &  186$^a$ &  L87 \\
 148184  & B2Vne  & 52481.1797 & $-$4.7$\pm$2.1 [34] &$-$19.0 &       &  148$^b$  &SB8,L87\\
 149438  & B0V    & 52414.3446 &    1.6$\pm$0.8 [43] &    1.7 &  0.8  &   10$^b$  &  L87 \\
\multicolumn{8}{c}{Other} \\
 104841  & B2IV   & 52411.0594 & $-$8.9$\pm$0.5 [73] &   16.1 &     I &   25$^b$  &  SB8 \\
 105435  & B2IVne & 52411.0725 &    3.8$\pm$2.8 [15] &   11.0 &     C &  298$^b$  & VDH\\
 105937  & B3V    & 52413.0558 &   22.7$\pm$1.5 [33] &   15.0 &     C &  129$^b$  &    \\
 109026  & B5V    & 52411.1069 &    4.0$\pm$1.6 [31] &    2.5 &     D &  188$^b$  & \\
 110335  & B6IVe  & 52481.0996 &    4.8$\pm$1.8 [23] &   12.5 &     B &  250$^a$ &   \\
 111123  & B0.5IV & 52411.1357 &    9.8$\pm$0.7 [53] &   10.3 &     A &   40$^b$  & VDH\\
 115846  & B3IV   & 52481.97   &$-$21.8$\pm$1.6 [27] &    3.0 &   4.0 &  168$^b$  &   \\
 116072  & B2.5Vn & 52411.1714 &   18.8$\pm$2.5 [21] &    3.0 &     C &  233$^b$  & VDH\\
 118716  & B1III  & 52413.1107 &   14.0$\pm$1.2 [32] &    3.0 &     B &  114$^b$  & VDH\\
 120640  & B2Vp   & 52413.1209 & $-$2.1$\pm$0.8 [63] & $-$4.7 &   0.8 &   21$^b$  &     \\
         &        & 52414.1219 & $-$1.5$\pm$0.6 [49] &        &       &       &    \\
 126341  & B2IV   & 52414.1712 &$-$21.1$\pm$1.0 [73] &$-$21.5 &    B  &   15$^b$  & VDH \\
 132058  & B2III  & 52414.2050 &    0.1$\pm$1.0 [48] &    0.2 &   0.9 &   92$^b$  & L87 \\
 132955  & B3V    & 52414.2424 &    5.1$\pm$0.5 [70] &    3.7 &   2.1 &    8$^b$  & VDH \\
 144218  & B2V    & 52414.3716 &    0.6$\pm$0.8 [57] & $-$5.6 &   0.8 &   56$^b$  &VDH,L87,ST,SB8\\
 151985  & B2IV   & 52414.3352 &    1.9$\pm$0.7 [56] &    1.3 &  0.7  &   52$^b$  & L87,SB8 \\
\hline
\end{tabular}
\end{center}
\begin{list}{}{}
\item * A: errors $\le$ 2.5 ${\rm km\,s}^{-1}$; B: 2.5 $<$ errors $\le$ 5.0 ${\rm km\,s}^{-1}$;
C: 5.0 $<$ errors $\le$ 10.0 ${\rm km\,s}^{-1}$;
D: errors $\ge$ 10 ${\rm km\,s}^{-1}$;
E: too uncertain  (from Table 3 of Barbier-Brossat \& Figon \cite{bbf00})
\item a: from Glebocki \& Stawikowski (\cite{gle00}) and
b: from Brown \& Verschueren (\cite{ver97})
\item SB8 - Eighth Orbital Elements of Spectroscopic Binaries (Batten et al. \cite{bat89})
\item L87 - Levato et al. (1987)
\item VDH - Visual Double Stars in Hipparcos (Dommanget \& Nys \cite{dom00})
\item ST - Shatsky \& Tokovinin (2002)
\end{list}
\end{table}


\begin{figure}
\centering
\includegraphics[angle=270,width=\textwidth]{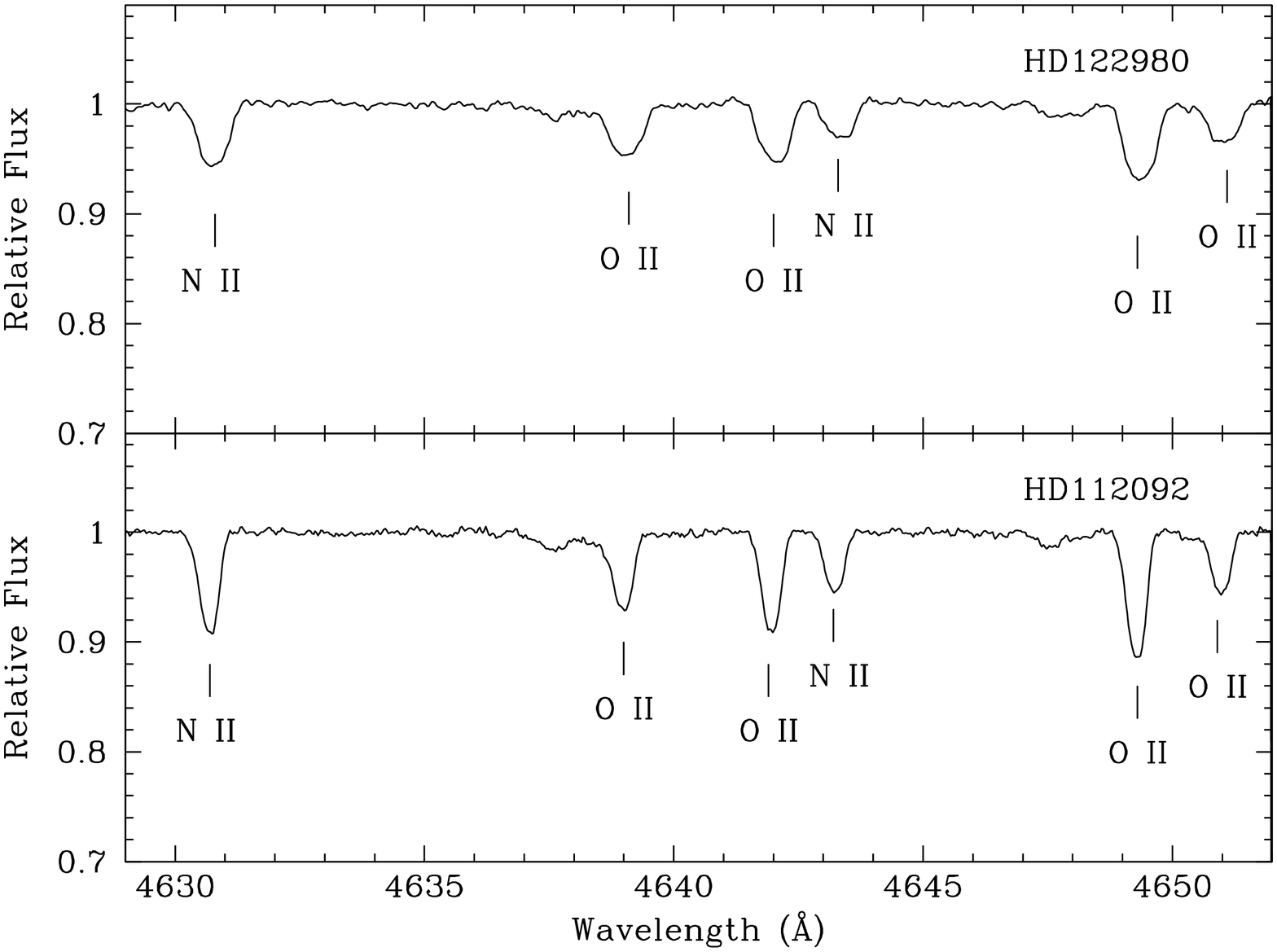}
\caption{Sample spectra for two target stars. The top panel shows
HD122980 and the bottom panel shows HD112092. The lines appearing
in this spectral region are identified. } \label{spectra}
\end{figure}

\begin{figure}
\centering
\includegraphics[angle=0,width=\textwidth]{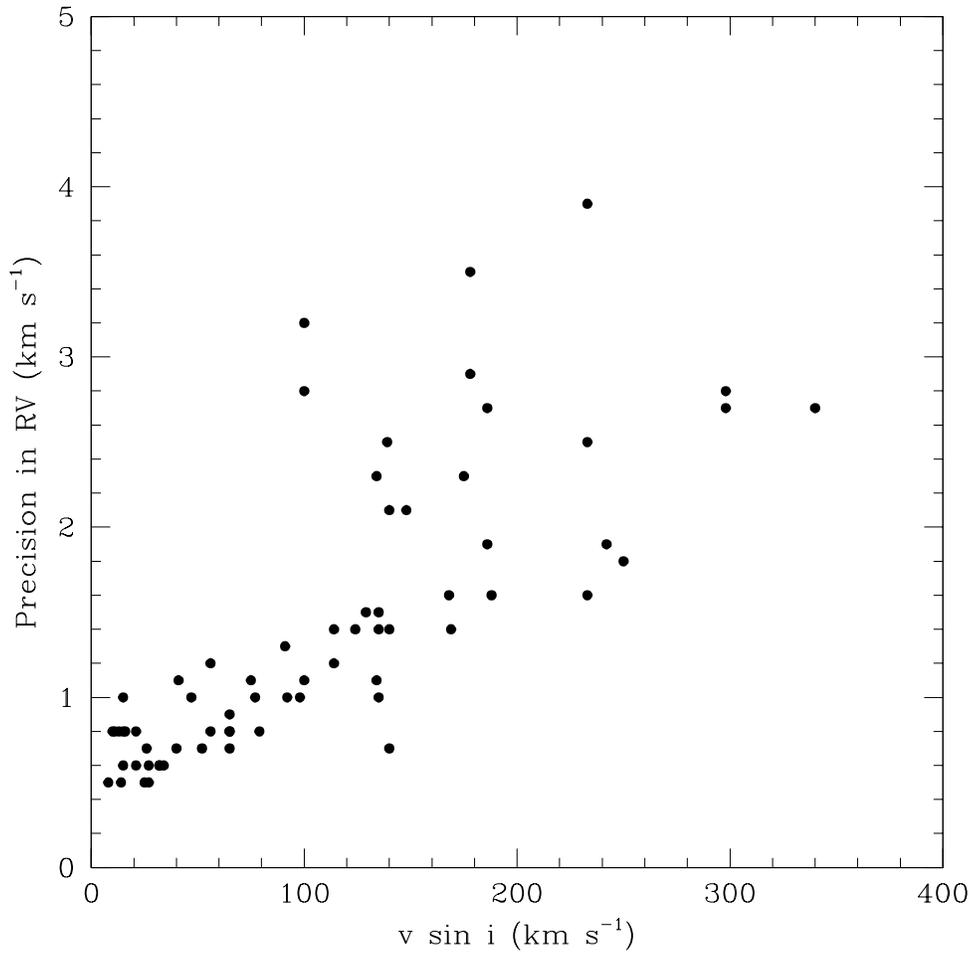}
\caption{The dependance of the internal r.m.s. errors of our
radial velocity determinations on the star's projected rotational
velocities ($v \sin i$) taken from Brown \& Verschueren (1997) and
Glebocki \& Stawikowski (2000).} \label{vrvseni}
\end{figure}

\begin{figure}
\centering
\includegraphics[angle=0,width=\textwidth]{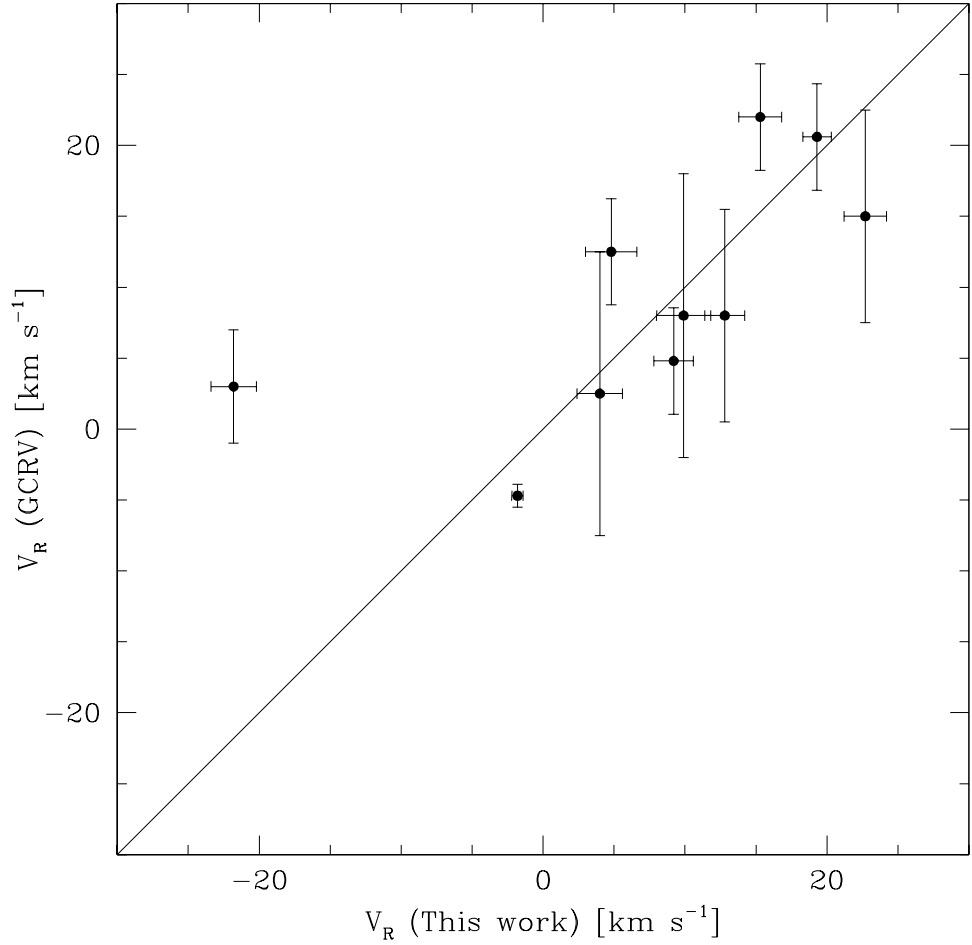}
\caption{A Comparison between the radial velocities derived in
this study with previously determined radial velocities from the
GCRV (Barbier-Brossat \& Figon 2000). The targets shown are those
stars from our sample considered to be single stars. The adopted
error bars for this study are the sigmas listed in column 4 of
Table 1. The adopted error bars from the GCRV were calculated as
the mean value of the range of uncertainties in RV corresponding
to the quality flags A-D (listed in Table~1). For those stars with
RV quality A and D we adopted uncertainties of 2.5 ${\rm
km\,s}^{-1}$ and 10 ${\rm km\,s}^{-1}$, respectively. The x=y line
representing perfect agreement is shown for comparison. The very
discrepant point at (-21.8,3.0) represents the target star
HD115846, which could be an unsuspected binary. } \label{vlit}
\end{figure}

\begin{figure}
\centering
\includegraphics[angle=270,width=\textwidth]{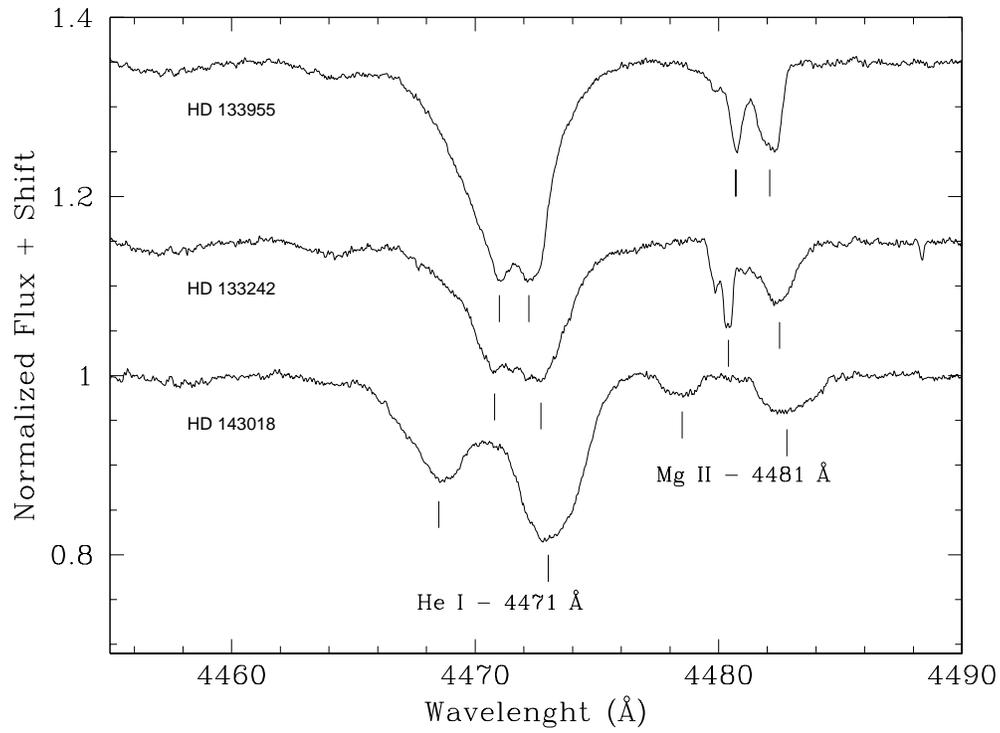}
\caption{Sample spectra of the double lined spectroscopic binaries
HD133242, HD133955 and HD143018 showing the lines 4471\AA \ of He I
and 4481\AA \ of Mg II.}
\label{sb2}
\end{figure}

\end{document}